\documentclass[aps,prl,twocolumn,superscriptaddress,amsfont,graphicx,preprintnumbers]{revtex4-1}
\usepackage{color,graphicx,epsfig}
\usepackage{ifpdf}
\usepackage{amsmath}
\usepackage{bm}
\usepackage{color}
\usepackage[english]{babel}
\usepackage{booktabs}
\usepackage{graphicx}
\usepackage{amsfonts}
\usepackage{amssymb}
\usepackage{hyperref}
\usepackage{slashed}
\usepackage{url}
\usepackage{array}
\usepackage{enumerate}

\definecolor{nicered}{rgb}{0.7,0.1,0.1}
\definecolor{nicegreen}{rgb}{0.1,0.5,0.1}
\hypersetup{colorlinks,citecolor= nicegreen,linkcolor= nicered}

\newcommand{\xgb}{\texttt{XGBoost}~}
\newcommand{\lgbm}{\texttt{LightGBM}~}
\newcommand{\openl}{\texttt{OpenLoops}~}
\newcommand{\tenr}{{\it`10~regions'}~}
\newcommand{\oner}{{\it`1~region'}~}
\newcommand{\msq}{\langle\left|\mathcal{M}\right|^2\rangle}
\newcommand{\cth}{\cos\theta}
\newcommand{\mz}{m_Z}
\newcommand{\ptx}{p_{T,\text{cut}}}

\definecolor{Red}{rgb}{1.,0.,0.}
\definecolor{Blu}{rgb}{0.1,0.3,0.9}

\graphicspath{{./figs/}}
\begin{document}
\def\DESY{DESY, Notkestra{\ss}e 85, 22607 Hamburg, Germany}
\def\PSI{Paul Scherrer Institut, Forschungsstra{\ss}e 111, 5232 Villigen PSI, Swizerland}
\def\Zurich{Physik-Institut,  Universit{\"a}t  Z{\"u}rich,  Winterthurerstrasse  190,  CH-8057  Z{\"u}rich,  Switzerland}
\preprint{DESY 19-232}
\title{\boldmath Machine Learning amplitudes for faster event generation}
\author{Fady Bishara}
\email[Electronic address:]{fady.bishara@desy.de}
\affiliation{\DESY}
\author{Marc Montull} 
\email[Electronic address:]{marc.montull@gmail.com}
\affiliation{\DESY}\affiliation{\PSI}\affiliation{\Zurich}
\begin{abstract} 
We propose to replace the exact amplitudes used in MC event generators for trained Machine Learning regressors, with the aim of speeding up the evaluation of {\it slow} amplitudes. As a proof of concept, we study the process $gg \to ZZ$ whose LO amplitude is loop induced. We show that gradient boosting machines like $\texttt{XGBoost}$ can predict the fully differential distributions with errors below $0.1 \%$, and with prediction times $\mathcal{O}(10^3)$ faster than the evaluation of the exact function. This is achieved with training times $\sim 7$ minutes and regressors of size $\lesssim 30$~Mb. These results suggest a possible new avenue to speed up MC event generators.
\end{abstract}
\maketitle

\section{Introduction}
The success of the LHC in discovering the Higgs boson is a testament to the impressive advancements made by the HEP community in understanding accelerators, detectors, and to make accurate Standard Model (SM) predictions. 
As a result, the LHC is rapidly evolving from an {\it energy frontier machine}, capable of discovering new resonances, to a {\it precision machine}, capable of measuring small deviations over precise SM predictions.

Due to the key role of higher order corrections in precision physics, there has been an Herculean effort in recent years to compute, store, and automate higher loop calculations for SM and Beyond the SM (BSM) predictions \cite{Alwall:2014hca,Hirschi:2015iia,Boughezal:2016wmq,Campbell:2019dru, Nagy:2003tz,Nagy:2001fj,Gavin:2010az,Camarda:2019zyx,Catani:2009sm,Catani:2015vma,deFlorian:2011xf,Actis:2016mpe,Hirschi:2011pa,Li:2015foa, Actis:2016mpe,Naterop:2019xaf,Peraro:2019svx,Camarda:2019zyx,Bern:2013zja, Bertone:2016lga, Heinrich:2016jad,Spira:2016ztx}.
As impressive as this has been, the use of N(N)LO results by the broader HEP community has been relatively low, in part due to the long times required to evaluate amplitudes beyond tree level.
This evaluation time increases dramatically with the loop order, and makes certain Monte Carlo (MC) event simulations at one loop already unfeasible.
Nonetheless, higher loop effects will become more important as the precision from the experimental and theoretical sides keeps improving.
This calls for innovations to reduce evaluation times for {\it slow} amplitudes.
One possible avenue to do just that is to improve the traditional tools and techniques -- an effort that is well under way.
In  this  work, however, we take a new and different approach to address these issues.

\medskip

{\bf The main goal of this work} is to show that thanks to the advances in Machine Learning (ML) algorithms and tools, it is now possible to train ML regressors with pre-computed {\it slow} amplitudes, and use them to predict the same amplitudes accurately and in a fraction of the time. 

\begin{figure}[b]
    \centering
    \includegraphics[width=0.8\linewidth]{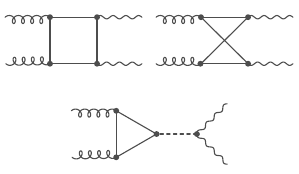}  
    \caption{SM LO diagrams for $gg \to ZZ$, up to fermion momentum flow and crossings.}
    \label{fig:box_diagram}
\end{figure}

As a proof of concept we study the $gg \to ZZ$ process which is loop induced at LO (see Fig.~\ref{fig:box_diagram}). We find that ML regressors can achieve prediction times $\mathcal{O}(10^3)$ faster than traditional tools while the predicted values for single and double differential distributions have errors below $0.1 \%$.
This was achieved with training times $\lesssim$ 7
minutes on a single CPU core, and with a disk size for the trained regressors of a few to tens of megabytes (Mb).

Machine Learning algorithms are constantly finding new applications in HEP research (see \cite{Bellagente:2019uyp,Andreassen:2019nnm,Nachman:2019dol,Bendavid:2017zhk,Otten:2019hhl,Klimek:2018mza,Bothmann:2018trh,Ball:2017nwa,Ball:2014uwa,Santos:2016kno, Farina:2018fyg,DAgnolo:2018cun, Blance:2019ibf,Cerri:2018anq,Brochet:2018pqf,Banerjee:2019twi,Komiske:2016rsd,Coccaro:2019lgs,Andreassen:2018apy,Andreassen:2019txo,1511.05190,1708.02949, 1805.02664,1902.02634,1911.09107,1705.02355,1712.10321,Chakraborty:2019imr,Lim:2018toa} for concrete applications and \cite{Luisoni:2016xkv, Buckley:2019wov,Abdughani:2019wuv,Albertsson:2018maf,Alves:2017she,Carrazza:2017qro,Larkoski:2017jix,Bourilkov:2019yoi,Brehmer:2019xox,Brehmer:2019gmn,Carleo:2019ptp} for recent reviews). Nonetheless, we are not aware of any work where these tools have been used to speed up time consuming amplitudes \footnote{The only similar works we found is Ref.~\cite{Otten:2018kum} where DNN are used to predict NLO cross sections for the pMSSM-19 and 
Ref.~\cite{Chawdhry:2019bji} ML regressors were used as interpolators to store the NNLO QCD amplitude for $p p \to 3 \gamma$.}.

There are many processes where speeding up MC event generation can be immediately useful. Therefore, a next step after this work would be to test the ML regressors on other processes and implement them into a MC generator. We comment on further applications at the end of this letter.
\section{A proof of concept with $gg \to ZZ$}
\label{sec:proofofconcept}
We chose to test the performance of ML regressors in approximating the  $gg \to ZZ$ squared amplitude for several  reasons. First of all, the LO contribution to this process arises at one loop (Fig. \ref{fig:box_diagram}) and therefore is {\it slow}.
Secondly, it was shown in Ref.  \cite{Cascioli:2014yka} that this process contributes the bulk ($~\sim 60\%$) of the full NNLO correction of hadronic Z-boson pair production, making its computation imperative when performing phenomenlogical studies to test the SM or to search for New Physics (NP).
In addition, it is relevant for NP searches where it constitutes a background to $p p \to ZH$ with $H$ decaying to $\bar{b}b$ or to invisible new particles \cite{Khachatryan:2016whc,Aaboud:2018zhk,Sirunyan:2018kst,Banerjee:2018bio, Banerjee:2019pks,Arcadi:2019lka}.
At the same time, this process is simple enough to avoid unnecessary complications: the squared amplitude only depends on two variables, the center of mass energy and the polar angle $\theta$, i.e. $|\mathcal{M}(\sqrt{\hat{s}}, \cos \theta)|^2$. Furthermore, when the the pair of $ZZ$ bosons are on-shell, there are no resonant peaks.
We leave the study of processes with $s$-channel resonances for future work.
Furthermore, since the $\alpha_s$ dependence amounts to an overall rescaling of the amplitude squared, we can approximate the function using a fixed value of $\alpha_s$ and restore the scale dependence afterwards.
\section{ML Algorithm and training}
\label{sec:TrainNPred}
\subsection{Choosing a Machine Learning algorithm}
\label{sec:ml_algo}
The problem we are trying to address here requires, above all, two features from an ML algorithm: first, it must be able to approximate the true function over the entire domain as accurately as possible; second, it must be be able to do so faster than existing dedicated programs $ \sim 5 \cdot 10^{-3}\;[s]$ per phase space (PS) point~\footnote{We determined this evaluation time using \openl~\cite{Buccioni:2019sur} to generate the exact $gg \to ZZ$ squared amplitudes for the training and prediction sets. The CPU core used for this timing is the same one used for all timings in this letter, see main text.}.
An additional bonus feature is for the model to be lightweight, i.e. to have a small disk size, $\lesssim \mathcal{O}(100)$ Mb, so that it is easy to distribute quickly.

With this in mind, we evaluated several algorithms suited for regression in the early stages of this work. In particular, we tested deep neural networks (DNN) with \texttt{TensorFlow} \cite{tensorflow2015-whitepaper}, random forests~\cite{ranger:2017,Ho:1995:RDF:844379.844681,Breiman2001}, and gradient boosting machines (GBM)~\cite{breiman1997arcing,friedman2001}.
From the outset, GBMs as implemented in \xgb \cite{Chen_2016}
outperformed the others by far in terms of speed, accuracy, and robustness against overfitting with very little tuning~\footnote{\xgb rose to prominence by winning the Kaggle Higgs Callenge. Since then it has been consistently on the top of the ladder in a large number of the Kaggle competitions \cite{KaggleRank1,KaggleRank2} outperforming other ML architectures, like DNN's or SVM's.}.
Therefore, all the results presented in this letter were obtained with \xgb via the \texttt{scikit-learn} API.
\medskip

As discussed above, we use the default or close to the default values for the hyper-parameters, except for the number of estimators ($n$), maximum depth of the trees ($m_d$), and the learning rate ($l_r$), for which we performed a small scan $n \in [10,1000]$, $m_d \in [10 , 800]$, and $l_r\in[0.01,0.3]$. Based on this bare bones optimization, the final set of parameters used in this work are given in Table~\ref{tab:xgb-machine}.

\begin{table}[t]
\begin{small} \begin{center}
\begin{tabular}{>{\raggedleft}p{0.3\linewidth}>{\centering\arraybackslash}p{0.4\linewidth}}
\toprule[1pt]
\xgb parameter          \quad    & \quad Value  \\
\midrule[0.5pt]
n\_estimators & 200 \\
max\_depth & 50 \\
learning rate & 0.1\\
\midrule[0.3pt]
min\_child\_weight & 1 \\
$\gamma$ & 0 \\
colsample\_bytree & 1\\
subsample & 0.75 \\
booster & gbtree \\
objective & reg:squarederror\\
\bottomrule[1pt]
\end{tabular}
\end{center} \end{small}
\caption{Hyper-parameter settings used for all \xgb regressors in this work. The parameters we attempted to optimize are shown above the split while the values for the parameters below the split are the \xgb default ones with the exception of `subsample', see text for details.}
\label{tab:xgb-machine}
\end{table}

\subsection{Datasets for Training and Prediction}
\label{sec:datasets}
To train and test the \xgb regressor, we generated $18$ million (18M) pairs of phase space points, $(\sqrt{\hat{s}},\cos \theta)$, uniformly distributed in the region defined by,
\begin{equation}
\begin{split}
\sqrt{\hat{s}} &\in \left[2\sqrt{\mz^2+\ptx^2}, 3 \, \mathrm{TeV}\right] \,,\\
\cth &\in [-1,1]\times \sqrt{1-\frac{4p^2_{T,\text{cut}}}{\hat{s}-4\mz^2}}\,,
\label{eq:PSregion}
\end{split}
\end{equation}
with $p_T^\text{cut} = 1$ GeV to regulate the singularity in $\msq$ in the limit $p_T \rightarrow 0$ (similarly to what is done in \texttt{MCFM}~\cite{Campbell:2010ff} and \texttt{Madgraph\_aMC@NLO}~\cite{Hirschi:2015iia}).
We chose  $(\sqrt{\hat{s}} \, )_{max}$ = 3 TeV as an arbitrary cutoff relevant for LHC physics. Nevertheless, it is straightforward, and inconsequential, to extend the cutoff to the collider center of mass energy; we checked this explicitly up to 14 TeV.
\medskip

We then computed the corresponding squared amplitudes required for training and testing \xgb, using the \openl software~\footnote{We also tested \texttt{MadLoops}~\cite{Hirschi:2015iia} and found it to be a factor of a few slower than \openl~\cite{Buccioni:2019sur} for the $gg\to ZZ$ process. For this reason, we used the faster evaluation time of \openl as a benchmark.}.
The full sample of $18$M points was split into a training and prediction dataset with 3M and 15M points, respectively.
To ensure that the data sets are statistically independent, we generated the phase space points using the \texttt{Python} implementation of the Mersenne Twister algorithm~\cite{Matsumoto:1998:MTE:272991.272995} which, when initialized properly, has a period of $2^{19937}-1$.

\subsection{Phase Space partitioning and multiple regressors}
\label{sec:oner_tenr}
The function that we are trying to approximate, $\msq$, is peaked at $\cth \to \pm 1$. This motivates an ansatz to break up the full phase space into smaller sub-regions with roughly equal integrated $d\msq/d \mbox{PS}$ with the purpose of training one regressor per sub-region. 

For example, choosing $\cth$ and $\sqrt{\hat{s}}$ regions defined by
\begin{gather}
    \cth\in\{-1,-0.94,-0.7,0.7,0.94,1\}\times \sqrt{1-\frac{4\ptx^2}{\hat{s}-4\mz^2}}\,, \nonumber \\
    \sqrt{\hat{s}}\in\{2\sqrt{\mz^2+\ptx^2},1.3\;\mbox{[TeV]},3\mbox{[TeV]}\}\,,
    \label{eq:PSregions}
\end{gather}
with the idea of decreasing as much as possible the variation of the squared amplitude in each sub-region.
This partitions the full phase space into 10 sub-regions each with its own dedicated \xgb regressor that is trained on, and predicts in, only one sub-region. These partitions are delineated by dashed gray lines in the right half of Fig.~\ref{fig:double_diff_distribution}.
\medskip

For the remainder of this letter, we will refer to the ansatz with 10 regressors as the {\it`10 regions'} regressor, and to the one trained on the full domain defined by Eq.~\eqref{eq:PSregion} as \oner. 
\medskip

To compare the performance of the \oner and \tenr regressors, we first train the \oner regressor on a given dataset. Then, to train each of the ten regressors that make up the \tenr regressor, we split the same dataset according to the regions defined in Eq.~\eqref{eq:PSregions}. In the end, each of these ten regressors making the \tenr regressor is only trained on a fraction between 2.4\% and 24\% of the total dataset, corresponding to the fraction of its phase space area (since the PS is uniformly sampled).
\subsection{Training time}
\label{sec:train_time}
\begin{figure}[t]
   \centering
  \includegraphics[scale=1]{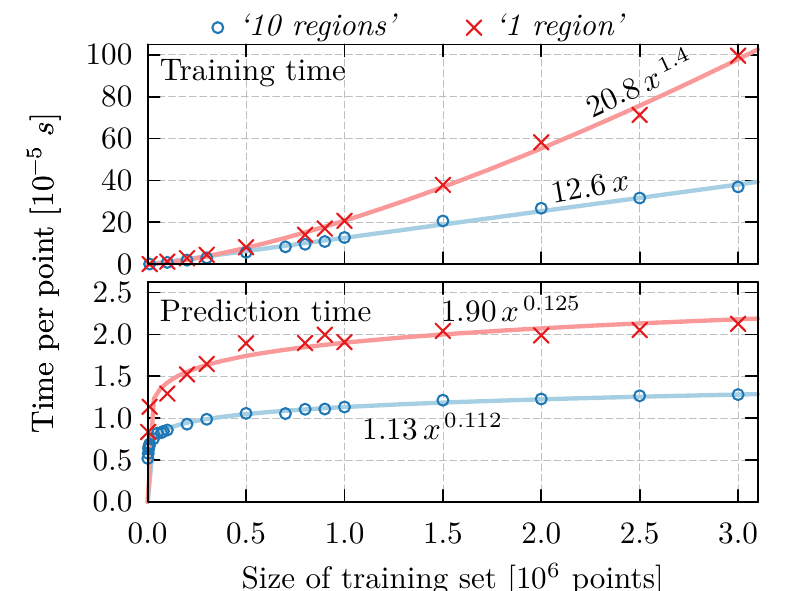}
 \caption{Prediction time per point as a function of the size of training set. The crosses (open circles) correspond to the results for the \oner (\tenr) regressors. The solid curves are simple power law fits and are shown in the legend.}
\label{fig:timing}
\end{figure}
We benchmark the time it took to train the \oner and each of the \tenr regressors on a single CPU core of an Intel\textsuperscript{\textregistered} Xeon\textsuperscript{\textregistered} CPU model \texttt{E5-2640V4} @ 2.40 GHz on \texttt{x86\_64} architecture.
Since \xgb can train and predict on multiple cores by default, the times reported here are quite conservative.  In practice, modern desktop machines with at least four cores are increasingly common and so training and prediction times can be easily be improved by a factor of a few to ten.
\medskip

The results of the timing tests for the training phase are shown in Fig.~\ref{fig:timing} (top panel) for both the \oner and \tenr regressors; in addition, a simple power law fit to the points is given.
For a training dataset size of 3M PS points, the \oner (\tenr) regressors took $\sim 16\,(7)$ minutes to train. In the case of the \tenr regressors -- there are 10 of them -- we added up the times it took to train each one of them.
%
\section{Results}
\label{sec:results}
\begin{figure}[t]
    \centering
    \includegraphics[scale=1]{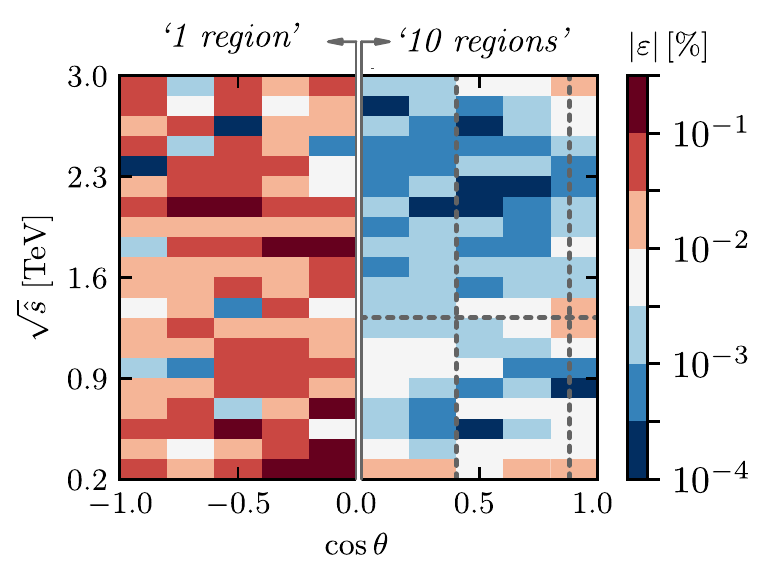}
    \caption{Absolute value of the percentage relative error per bin of the double differential distribution, $d^2\msq/d\cos \theta d\sqrt{\hat{s}}$. Each bin has size $140 \times 0.2$ (GeV, $\cos \theta$). The total number of training (prediction) points is 3M (16M). {\bf Left:} \oner regressor. {\bf Right:} \tenr regressor with dashed gray lines showing the sub-regions defined in  Eq.~\eqref{eq:PSregions}.
    }
    \label{fig:double_diff_distribution}
\end{figure}

In order to benchmark the trained \oner and \tenr regressors defined above, we study the relative error of their predictions, and measure their evaluation times. The relative error is defined as,
\begin{equation}
\label{eq.error}
\varepsilon=\frac{\msq_{\mathtt{OpenLoops}}-\msq_{\mathtt{XGBoost}}}{\msq_{\mathtt{OpenLoops}}} \,.
\end{equation}
As for the training times, we measure the prediction times on a single core of the same CPU described above.
\subsection{Accuracy of predictions}
\label{sec:accNspeed}
Figure~\ref{fig:double_diff_distribution} shows the relative error on the sum of the predicted $d^2 \msq / d\mbox{PS}$ values in each bin for the \oner and \tenr regressors.
Each of the regressors was trained on 3M points and the relative errors were computed from predictions of 15M points.
Each bin has a size of 140 GeV $\times\,0.2$, which is appropriate for phenomenological studies at the LHC.
The left and right sides of the plot correspond to the \oner and \tenr regressors, respectively, put together for easier comparison since the amplitude is symmetric under $\cos \theta \to - \cos \theta$.
In addition, the right panel is overlaid with the boundaries of the sub-regions defined in Eq.~\eqref{eq:PSregions}.
We find that the \oner regressor has a maximum relative error per bin of $0.3\%$ while the \tenr regressor has a maximum error of $0.03\%$.
\medskip

\begin{figure}[t]\centering
	\includegraphics[scale=1]{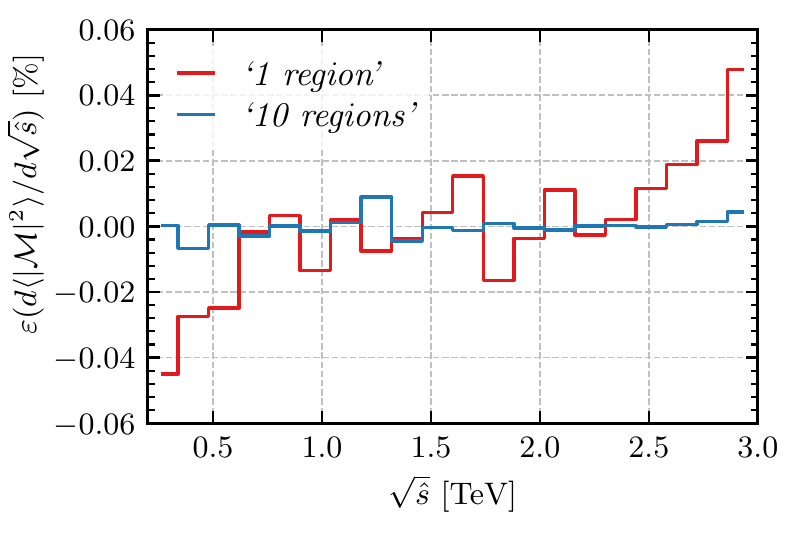}
	\caption{Relative error of $d \msq / d\sqrt{\hat{s}}$. The red and the blue curves correspond to the \oner and \tenr regressors respectively (see Sec. \ref{sec:oner_tenr}). }
	\label{fig:single_diff_dist}
\end{figure}
For phenomenological studies, another important differential distribution is the singly differential one with respect to $\sqrt{\hat{s}}$. The relative error for this distribution is shown in Fig.~\ref{fig:single_diff_dist} and is of $\mathcal{O}$(percent) and $\mathcal{O}$(permille) for the \oner and \tenr regions respectively.
\medskip

\begin{figure}[t]
	\centering
	\includegraphics[scale=1]{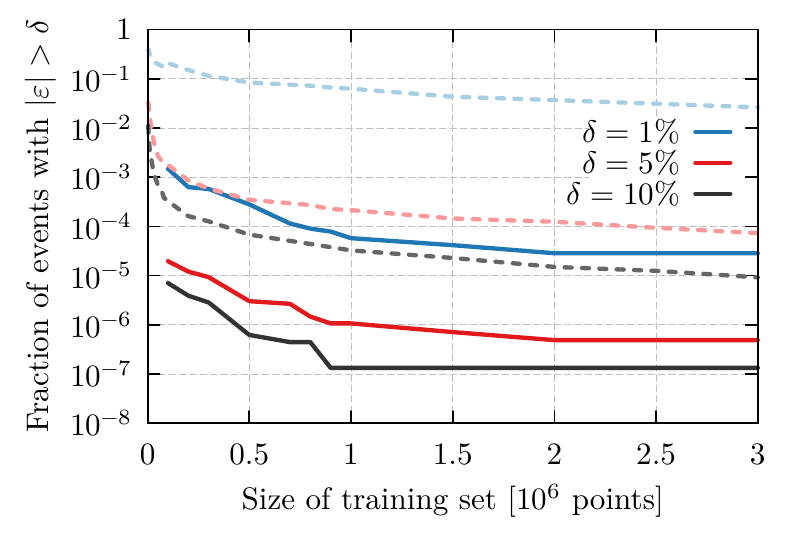}
	\caption{Percentage of predicted points with an error greater than $1\%$ (blue), $5\%$ (red), and $10\%$ (black) as a function of the number of trained points (the number of predicted points is 15 million).	The solid (dashed) curves correspond to the \tenr(\oner) regressors.}
	\label{fig:error_vs_npoints} 
\end{figure}

In order to assess the effect of the size of the training set on the performance of the machines, we show in Fig.~\ref{fig:error_vs_npoints} the fraction of points with relative error greater than 1\%, 5\%, and 10\% using the full 15M phase space point prediction dataset. The dashed (solid) curves correspond to the full (subdivided) phase space. Again, it is clear that subdividing the phase space is very effective in reducing the errors.
Figure~\ref{fig:error_vs_npoints} also shows that, for the chosen hyperparameters (Table~\ref{tab:xgb-machine}), there is little benefit from using training datasets larger than 1M points.
Furthermore, we find no over-training even up to training datasets of 3M points. 
\medskip

From the results presented in this section, the benefit of subdividing the phase space and training separate machines on the subregions is clear: the error between the \oner and \tenr is reduced by an order of magnitude, see Fig. \ref{fig:double_diff_distribution} and Fig.~\ref{fig:single_diff_dist}, while the training and prediction times are reduced by a factor of two, see Fig.~\ref{fig:timing}.
\subsection{Prediction speed}
The time required to predict one phase space point is a crucial performance metric for the trained machine. Clearly it must be much faster than the time required to evaluate the true function (we use \openl as our benchmark). The results of the timing tests for the training phase are shown in Fig.~\ref{fig:timing} for both the \oner and \tenr regressors. In addition, a simple power law fit to the points is shown for each one on the plot. For the \tenr regressor trained on 1M points, the prediction time is $\sim 1\times 10^{-5}$ seconds in comparison to $8.7\times 10^{-3}$ seconds for \texttt{Fortran} interface of \openl -- i.e., a factor of $\sim 1000$ speedup.

Note that the trained regressors can be packaged as a single, standalone, \texttt{C} library. We checked that calling this library during an event generation run has negligible overhead.
\subsection{Disk size}
Another desirable feature for the standalone packaged library is to be lightweight in terms of disk size.
We find that for 1M points, the \oner (\tenr) regressor has a size is 2.6 (19) Megabytes.
This makes these regressors ultra portable and could be downloaded on the fly during MC event generation -- we envision machines of this type to be an option given to the user when generating events with popular MC event generators such as \texttt{MG5\_aMC\@NLO}~\cite{Alwall:2014hca}, \texttt{Sherpa}~\cite{Bothmann:2019yzt}, and \texttt{Whizard}~\cite{Kilian:2018onl}.
\section{Summary and conclusions}
\label{sec:Conclusions}
The idea of using ML regressors to approximate squared amplitudes proposed in this work is a new application of Machine Learning techniques in HEP. Our goal is to accurately predict the trained squared-amplitudes in a fraction of the time it takes to evaluate the exact ones.

As a proof of concept, we studied the accuracy and speed of the \xgb regressor to predict the squared amplitudes for the $gg \to ZZ$ process which at LO is generated at one loop. Our results show that the \xgb regressors deliver a {\bf 1000}-fold  speedup in evaluation time with respect to \openl with no more than $\mathbf{0.03 \%}$ relative error with respect to the {\it true} double differential distribution binned as in Fig.~\ref{fig:double_diff_distribution}.
 
Another convenient feature of the \xgb regressor studied in this letter, is its reduced training speed. Using the hyper-parameters given Table~\ref{tab:xgb-machine}, training on 1M uniformly generated PS points takes about {\bf 2} minutes on one CPU core.
Moreover, since \xgb is by default able to train and predict on multi-core CPUs, actual training and prediction times will be in practice faster by a factor proportional to the number of available CPU cores.
\xgb can also run on GPUs with some minor modifications, otherwise, \lgbm \cite{NIPS2017_6907} works on GPUs by default and could even be a better performing regressor. 

In addition, the disk size of the trained \xgb regressor for this process, is at most $\mathbf{30\;\mbox{Mb}}$, making it easy to distribute on the fly during process generation in MC event generators.
\medskip

Another important result of this work is to demonstrate that the errors on the predictions of the \xgb regressor can be reduced by an order of magnitude by training independent regressors on separate sub-regions of the full phase space.
A bonus feature of training more regressors on sub-regions is that their aggregate training and prediction times for a given dataset are reduced with respect to training a single regressor on the full phase space
\medskip

In Table~\ref{tab:summary} we summarize the aforementioned performance benchmarks for one \xgb regressor (\oner) trained on the full phase space region, and for ten regressors (\tenr) each trained on a sub-region. 
\medskip

\begin{table}[h]
\centering\renewcommand\arraystretch{1.3}
\begin{tabular}{>{\raggedleft}p{0.3\columnwidth}>{\centering}p{0.24\linewidth} >{\centering}p{0.24\linewidth}>{\raggedright\arraybackslash}p{0.148\linewidth}}
\toprule[1pt]
&{\oner}& {\tenr}& \\\hline

$\left | \varepsilon^\text{bins}_{min} \right|~[\%]$ &  $7\cdot 10^{-5}$  & $3\cdot 10^{-5}$ & Fig.~\ref{fig:double_diff_distribution}\\
$\left| \varepsilon^\text{bins}_{max}\right|~[\%]$ &  $0.3$  & $0.03$ & Fig.~\ref{fig:double_diff_distribution}\\
$t_\textrm{predict}^\text{(1 core)}~[s/\textrm{point}]$  & $2\cdot 10^{-5}$ & $10^{-5}$ & Fig.~\ref{fig:timing}\\ 
$t_\mathrm{train}^\text{(1 core)}~[s]$ & 977 & 390 & Fig.~\ref{fig:timing}\\
$\mathrm{Size}\;[\mathrm{Mb}]$ & $4.8$ & 28 &\\
\bottomrule[1pt]
\end{tabular}
\caption{Main characteristics of the two ML regressors trained on 3M points and predicting on 15M.
The relative errors $\varepsilon^\text{bins}_{max}$ and $\varepsilon^\text{bins}_{min}$ stand for the relative errors in the bins of size $140~\times~0.2$ ($\sqrt{\hat{s}}$ [GeV] $\times \, \cos \theta$).}
\label{tab:summary}
\end{table}
The success of the proof of concept studied in this work suggests many applications and further ideas to explore:
\begin{itemize}
    \item To test the performance of ML regressors on qualitatively different channels with {\it slow} amplitudes. For instance: {\it i)} Amplitudes with resonant s-channels {\it ii)} N(N)LO amplitudes {\it iii)} $2 \to n$ processes.
	\item  Test and benchmark other ML algorithms.
	\item Implement the trained ML regressors into an MC event generator. 
	\item On a side note, it would be interesting to test the performance of GBM on interpolating PDFs and NNLO grids. 
\end{itemize}

\medskip

\begin{acknowledgments} 
\paragraph{\bf Acknowledgements.} We are especially thankful to P. Englert for his contributions during the initial stages of this project. We are happy to thank M. F. Zoller and J. Schlenk for useful discussions.
We would like to thank J. Zupan for encouraging us to think about new applications of Machine Learning in HEP.
F.B.~was supported by the ERC Starting Grant NewAve (638528).
M.M.~was supported by the Swiss National Science Foundation, under Project Nos.~PP00P2176884.
This research was partially supported by the Munich Institute for Astro- and Particle Physics (MIAPP) which is funded by the Deutsche Forschungsgemeinschaft (DFG, German Research Foundation) under Germany's Excellence Strategy -- EXC-2094 – 390783311,
the Aspen Center for Physics, which is supported by National Science Foundation grant PHY-1607611,
and the Deutsche Forschungsgemeinschaft (DFG, German Research Foundation) under Germany's Excellence Strategy – EXC 2121 ``Quantum Universe'' -- 390833306.
\end{acknowledgments}
%
\bibliography{paper}
\end{document}